\documentclass[conference]{acmsiggraph}

\usepackage{graphicx}
\usepackage[normalem]{ulem}

\newcommand{\featname}{Pyramid-of-Parts}

\TOGonlineid{}
\TOGvolume{0}
\TOGnumber{0}
\TOGarticleDOI{1111111.2222222}
\TOGprojectURL{}
\TOGvideoURL{}
\TOGdataURL{}
\TOGcodeURL{}

\title{Sketch-based Shape Retrieval using \featname}

\author{Changqing Zou$^{1,2}$ \thanks{aaronzou1125@gmail.com}\qquad Zhe Huang$^3$ \thanks{co-first author}\qquad Rynson W. H. Lau$^3$ \qquad Jianzhuang Liu$^2$ \qquad Hongbo Fu$^3$\\
 $^1$Hengyang Normal University \qquad $^2$Shenzhen Institutes of Advanced Technology, CAS \qquad $^3$City University of Hong Kong}

\pdfauthor{Changqing Zou, Zhe Huang, Rynson W. H. Lau, Jianzhuang Liu, and Hongbo Fu}


\begin{document}

 \teaser{
 	\centering
	\includegraphics[width=\linewidth]{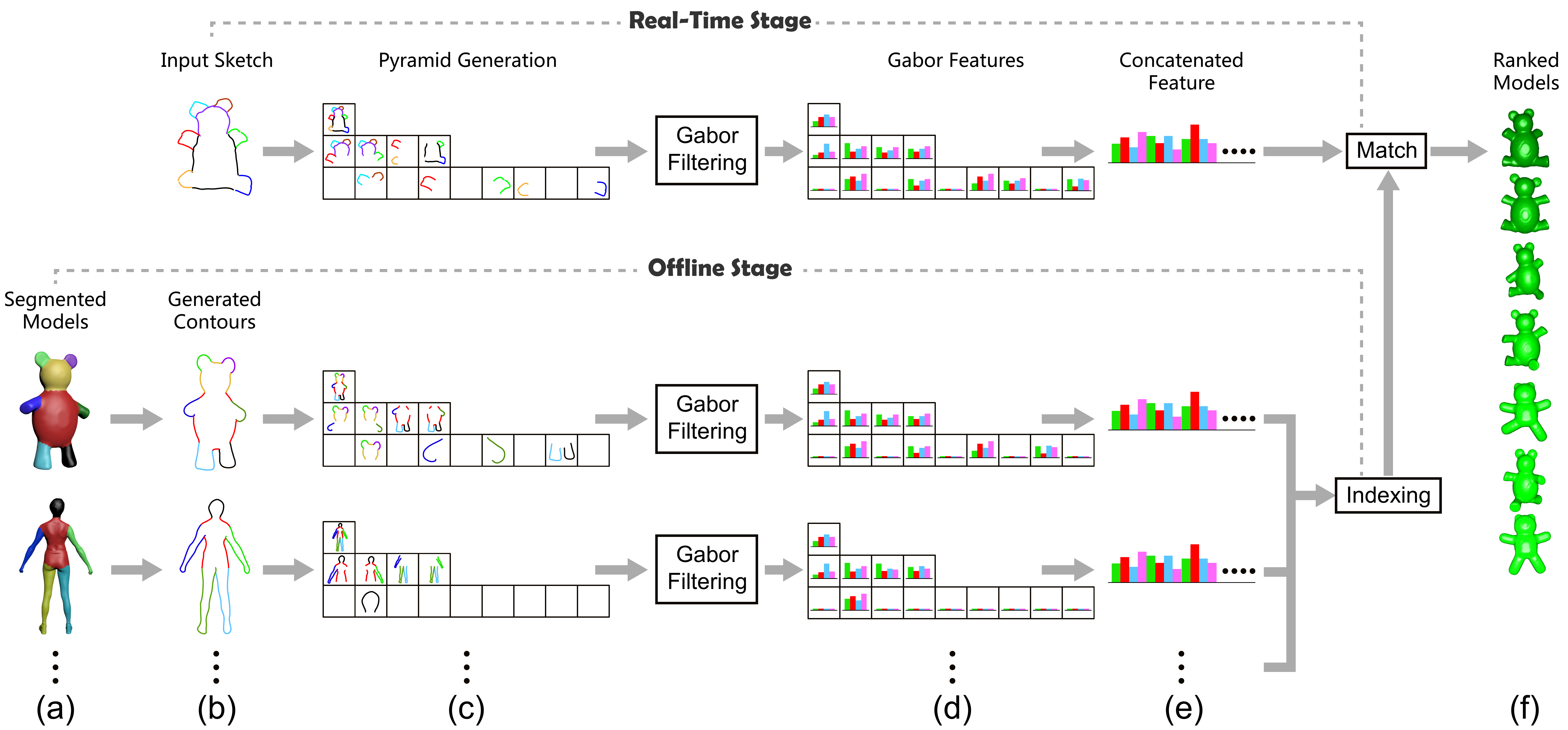}
	\caption{Our shape retrieval pipeline. The first row shows the pipeline for processing the input sketch. The input sketch is pre-segmented into semantic parts (color-coded) (b), which are assigned to different image regions and then grouped into a pyramid (c). Gabor features are then extracted from each group of parts (d), which are concatenated into the \emph{\featname}~feature (e). The last two rows illustrate the processing of database models. A pre-segmented 3D model (a) is first rendered into a 2D model contour using Suggestive Contours (b), where segmentation is transferred from the model. After that, the process is similar to processing of the input sketch. Finally, the input sketch is matched with each contour (or view) of every model, and a ranked list of similar models is returned (f).}
	\label{fig:pipeline}
 }

\maketitle

\begin{abstract}
	We present a multi-scale approach to sketch-based shape retrieval. It is based on a novel multi-scale shape descriptor called \emph{\featname}, which encodes the features and spatial relationship of the semantic parts of query sketches. The same descriptor can also be used to represent 2D projected views of 3D shapes, allowing effective matching of query sketches with 3D shapes across multiple scales.
%The proposed method relies on semantic segmentation of the query sketches as well as the 3D models, and it is possible that these segmentations are inconsistent. To cope with this problem, we devised a multi-scale grouping strategy that proves to be robust against these inconsistencies.
Experimental results show that the proposed method outperforms the state-of-the-art method, whether the sketch segmentation information is obtained manually or automatically by considering each stroke as a semantic part.
%	Our sketch-based shape retrieval system achieves the state-of-the-art accuracy, based on the evaluation over more than 600 sketches.
\end{abstract}

%\begin{CRcatlist}
%  \CRcat{I.3.3}{Computer Graphics}{Three-Dimensional Graphics and Realism}{Display Algorithms}
%  \CRcat{I.3.7}{Computer Graphics}{Three-Dimensional Graphics and Realism}{Radiosity};
%\end{CRcatlist}

\keywordlist

%% Use this only if you're preparing a technical paper to be published in the
%% ACM 'Transactions on Graphics' journal.

\TOGlinkslist

%% Required for all content.

\copyrightspace

\section{Introduction}

% background
Due to their simplicity and intuitiveness, sketch-based interfaces have been popular for 3D shape retrieval. A standard approach is to turn the 2D-3D matching problem involved into a 2D-2D matching problem by first rendering every 3D repository model as 2D contours under multiple views and then matching the query sketch with every resulting contour. How to define effective features to represent both input sketches and 2D contours is a key challenge in sketch-based shape retrieval. We use 2D contours of models and model sketches interchangeably in the following discussion.

\begin{figure}[h]
	\centering
	\includegraphics[width=\linewidth]{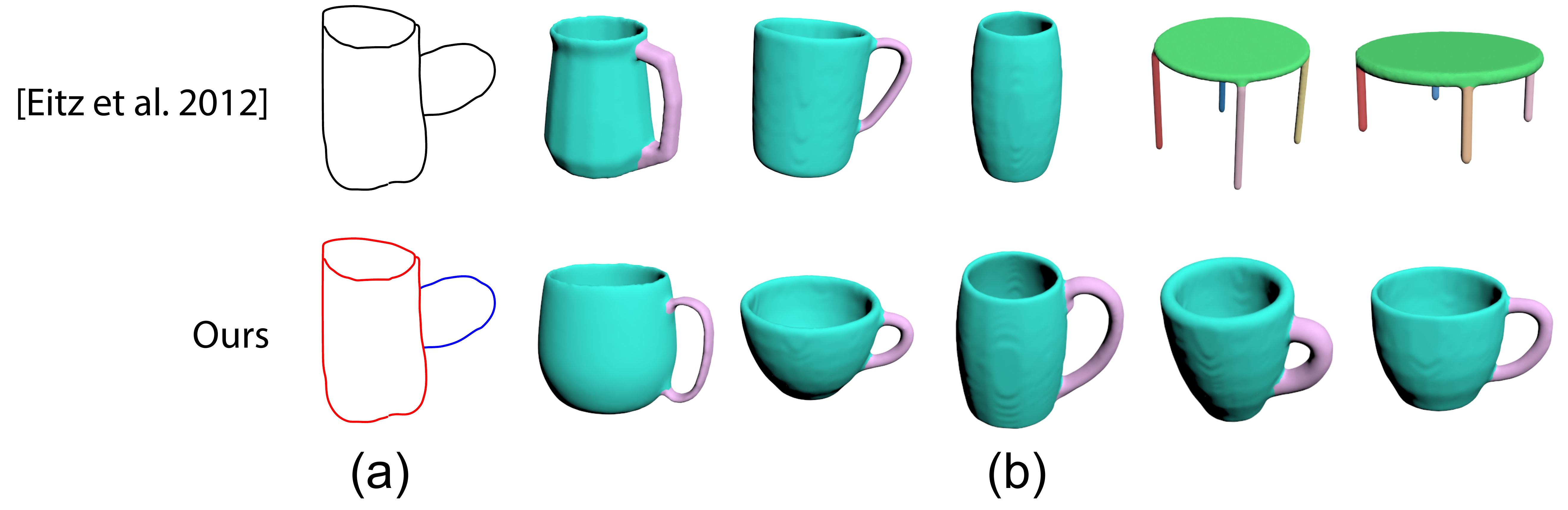}
	\caption{
Compared with \protect\cite{Eitz:2012:SketchRetrieval}, our sketch-based retrieval method based on the \emph{\featname}~returns more relevant models. The segmentation in the query sketch is color-coded.
%Top 5 models in (b) retrieved by~\protect\cite{eitz2012sketch} and by our method given the input sketches in (a).~\protect\cite{eitz2012sketch} uses bag-of-features model that discards spatial relationship among the features, and might return totally unrelated results (e.g., the last model in 3rd row). Our approach extracts features from semantic parts of the sketch and retains their spatial relationship, achieving better results. To enable such an approach, we require segmentation of the input sketch (a), as denoted by the different colors of strokes.
}
	\label{fig:motivation}
\end{figure}

% problems and % motivations
Various feature presentations (e.g., %Fourier Descriptors~\cite{Persoon:1986FD},
Gist~\cite{oliva2001modeling}, Spherical Harmonics~\cite{funkhouser2003search}, Eccentricity~\cite{li2012sketch}) have been proposed to
represent both queries and model sketches \emph{globally}. Such global sketch representations are able to encode high-level shape information, but sensitive to intra-class variation and shape deformation.
Recently, BoW representation~\cite{Eitz:2012:SketchRetrieval} has been brought into the field to address this problem. This representation is based on the statistics of the local features,  such as GALIF~\cite{Eitz:2012:SketchRetrieval} and SIFT~\cite{li2013shrec}, and it is proven more robust against the variations in the query and model sketches. However, approaches based on this representation may easily return locally similar but globally very different models (Figure~\ref{fig:motivation}). This is because the local features are still defined at the pixel level, without leveraging any high-level semantics, such as the semantically meaningful \emph{parts} in the sketch and the 3D models.
%\addtext{because the underlying representation is an aggregation of local information, and does not encode spatial information as well. The problems of the representations mainly utilizing global or local information about the sketch motivated us to represent sketches in a multi-scale manner.}

Part-level representations have been proven useful for object detection and recognition~\cite{FelzenszwalbH05} in the computer vision community. However, existing sketch representations are still largely defined at the pixel level only.
On the other hand, many techniques have been proposed to consistently decompose a set of 3D models into semantically meaningful parts~\cite{Kalogerakis:2010,Huang:2011,hu2012co}. In recent years, several techniques have also been developed to semantically segment freehand sketches, either automatically or interactively~\cite{Noris:2012a,Sun:sketchsegmentation2012,Huang:2014sketch}. Hence, it is interesting to explore if the use of semantic parts could lead to more discriminative sketch representations for retrieval.

In this paper, we present a new sketch representation, called \emph{\featname}, for sketch-based shape retrieval.
Our representation is derived from the available part-level information associated with the query sketch and the 3D repository models.
We consider two ways of obtaining sketch segmentation information, manually specified and automatically obtained simply by assuming that each input stroke forms a semantic part.
As the semantic segmentation of the query sketch and that of the 3D models might be different or at different levels of detail due to the multi-scale nature of objects, we thus adapt the idea of image pyramids to encode semantic parts in a multi-scale manner. Our retrieval algorithm will then compare the \emph{\featname}~of the input sketch with those of the 3D models across different scales and return the list of models ranked according to how well they match with the input sketch semantically.
% \hongbo{This paragraph might need to be extended a bit.}

%To address the above problems, we propose in this paper an all-encompassing shape representation, called \emph{\featname}, that combines the advantages of both low-level and high-level representations into a single effective representation.
%%as a compromise between low-level and high-level representations.
%%The same representation is used to model the input sketch as well as the projected line drawings of 3D models.
%Given an input sketch, our algorithm would automatically decompose it into coarse ``semantic parts" at multiple scales of the \featname. At the lowest level, these semantic parts may depict, for example, the ear of a bear. At a higher level, they may depict the upper body of the bear. (See Section~\ref{sec:partdef} for a precise definition of semantic parts.)
%The concept of ``part" is originated from research in object recognition, and has been widely explored in the relevant disciplines [add references].
%

\begin{figure}[tbh]
	\centering
	\includegraphics[width=\linewidth]{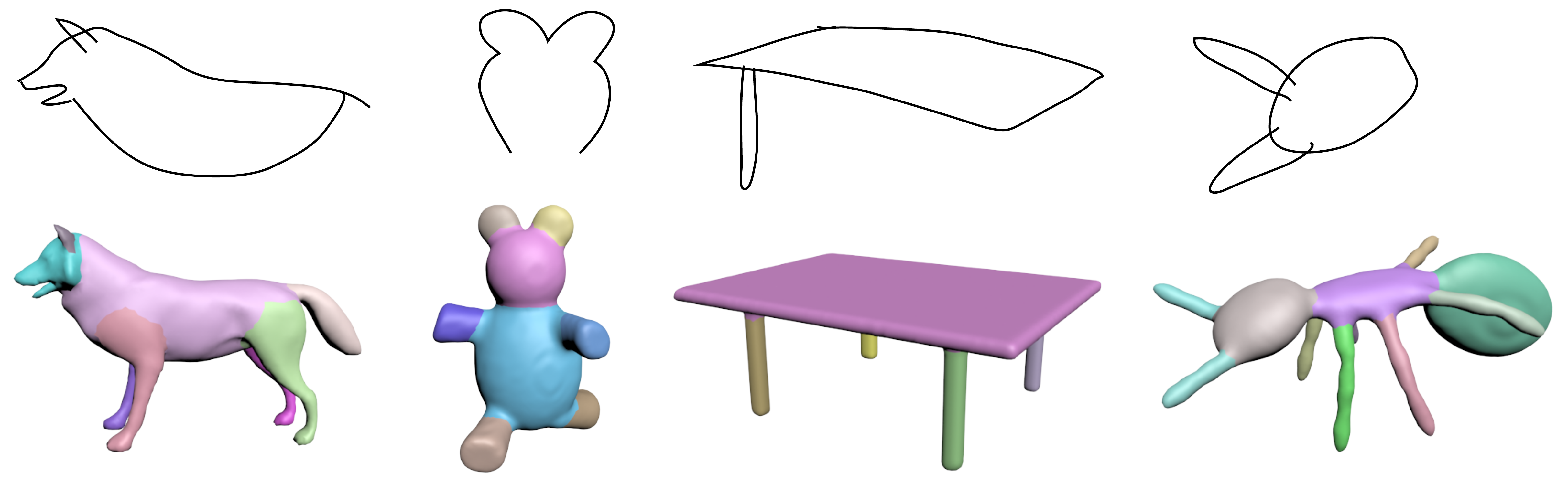}
	\caption{Given some incomplete query sketches (top row), our method is able to return the desired models (bottom row).}
	\label{fig:motivation2}
\end{figure}

Thanks to the \emph{\featname}, our sketch-based shape retrieval technique outperforms the state-of-the-art techniques, which are based on either local descriptor~\cite{Eitz:2012:SketchRetrieval} or global descriptor~\cite{zou2014VAR}, on two sketch datasets.
%derived from \cite{Chen:2009} and \cite{Eitz:2012:SketchRetrieval}\hongbo{please check}.
We present a new sketching interface that supports the commonly used coarse-to-fine drawing practice and naturally provides semantically segmented sketches as query sketches.
Since our representation encodes spatial information of the query sketch, our technique often produces desired results even if only a subset of parts are depicted in a query sketch (Figure~\ref{fig:motivation2}) and performs better than ~\cite{Eitz:2012:SketchRetrieval} in this task. We refer to this kind of matching based on incomplete input sketches as \emph{incomplete matching} in this paper.
%\hongbo{I'm not sure if this task should be called as partial matching. Any better term?} \james{constrained partial match?} \ryn{Rynson: how about incomplete matching? ***)}

%The rich discriminative information from the multi-scale \featname proves to be more effective than either the local or the global descriptor. In addition, our proposed representation is well suited for the task of partial matching, where the user only needs to draw part of the sketch (Figure~\ref{fig:motivation2}). This is because users typically stop drawing when some semantic parts are completed, which will get encoded in the proposed feature and compared with semantic parts of similar size and location in the database. This is more effective than comparing the partial sketch to the full sketch as in~\cite{eitz2012sketch}. We conducted experiments on two query modes -- complete matching and partial matching,  on PSB database \cite{Chen:2009}, and using the proposed representation, we achieve much better results.

%\begin{figure}[h]
%	\centering
%	\includegraphics[width=\linewidth]{naturalseg}
%	\caption{User strokes (a)(b) often imply functional segmentations, as seen by their similarity with (c)(d).}
%	\label{fig:naturalseg}
%\end{figure}

\section{Related Work}
%Currently, most of the popular representation methods are based on local features, which describe shape information of local patches around key-points. These key-points are sampled either via interesting point detectors or in a dense and uniform way.
%Feature descriptors such as GALIF \cite{Eitz:2012:SketchRetrieval} and HOG \cite{Eitz:2010cag}
%are extracted from the key-points and then clustered into a codebook.
%Given a query sketch, local features are first extracted and then quantified using the pre-learned codebook, which transforms the query sketch into the Bag-of-Features (BoF) representation. To a certain degree, this representation is able to discriminate common patterns among different sketchs/views, and is robust to intra-class variation as well as shape deformation. However, it suffers from the lack of spatial information in its encoding, and may erroneously consider unrelated shapes similar, such as the examples shown in Figure~\ref{fig:simplecompare}. Also, in~\cite{Eitz:2012:SketchRetrieval}, the local descriptors are extracted from blocks of the sketch, which might just contain random sets of strokes bearing no semantic information. We argue that extracting features from groups of strokes that resemble semantic parts of the sketch have more discriminative power, and gives better result.

%\hongbo{I completely rewrote this paragraph, since you're supposed to discuss how the existing works are relevant to ours instead of simply summarizing how each of them works.}\\

\textbf{Sketch-based Shape Retrieval.} This problem is often tackled by finding a repository model which rendered 2D contour (i.e., a silhouette rendering of the 3D model) under a certain viewpoint best matches the query sketch. Existing solutions mainly differ in the feature descriptors used to represent query/model sketches and can be largely categorized into two groups.
The first group of approaches make use of global descriptors (e.g., \cite{loffler2000content,funkhouser2003search}) to represent the sketch globally. However, global descriptors are sensitive to intra-class variations and shape deformation, which would bring global changes to the descriptors. Global descriptors have difficulty in handling incomplete query sketches, as the missing information would also affect the global descriptors.
% since because a moderate change in a part of the shape will affect all parts of the feature.

%\cite{loffler2000content} proposed one of the earliest global descriptor based methods, which accepts a user sketch to refine the initial keyword-based search result.
%~\cite{funkhouser2003search} described another 3D model search engine which supports sketch-based search. Global descriptors inspired by spherical harmonics are computed for the sketch and the 2D contours of the 3D models under different views, based on which the sketch is matched against all the contours. Their shape feature is rotational invariant and captures the coarse shapes well, but cannot differentiate fine details between shapes.
The second group of approaches use statistics about local descriptors for sketch representation. For example,
Yoon et al.~\shortcite{yoon2010sketch} represent sketches using statistics of their diffusion tensor fields, leading to a histogram of orientations.
Saavedra et al.~\shortcite{saavedra2012sketch} represent a sketch by the HOG feature of its ``key shape'', which is an approximation of the contour with straight lines.
Eitz et al. \shortcite{Eitz:2012:SketchRetrieval} adopt the Bag-of-Features (BoF) model to represent a sketch as a histogram of visual words.
%\ryn{These methods strike a balance between local and global features of 2D shapes (*** If they are local methods, how can they strike a balance between local and global methods? ***)}\james{Now we do not have the notion of ``local methods'' anymore, as this term is really confusing, perhaps ``local methods'' should refer to those methods that build point-to-point correspondence between sketches, and we might need to cite them}
These methods strike a balance between local and global features of 2D shapes, and are able to tolerate the inaccuracies inherent in sketches to some extent. However, since they discard spatial relationship among local descriptors, they often return unrelated shapes as similar (Figure~\ref{fig:motivation}). Such spatial information of local descriptors can be captured by our proposed \emph{\featname}, resulting in more discriminative power.
Moreover, there are some other methods (e.g., \cite{shao2011discriminative,li2013sketch}) that directly align the query sketch to
rendered model views to compute sketch-to-model distances. The main limitation of these approaches is that
they usually suffer from the efficiency problem.

\textbf{Sketch-based Image Retrieval.}
Sketch-based image retrieval has been under intensive research since 1990s, and many shape descriptors have been explored. (See~\cite{Eitz:2010cag} for a nice survey.) Among them, Histograms of Oriented Gradients (HOG)~\cite{Dalal:2005} and Shape Context~\cite{Belongie:2002} have become very popular due to their simplicity, generality and discriminative power. The BoF model can be built upon these local descriptors~\cite{eitz:2011}, and the resulting feature is shown to be more tolerant to sketch variations. These shape descriptors can be extended to multi-scale, leading to, for example, multi-scale HOG~\cite{Hu:2010} and multi-sample BoF~\cite{Zhong:2009}, where the sampled image patches do not have a fixed size. Compared to these works, our multi-scale descriptor is defined over semantic parts in the sketch, rather than image patches which content might bear no semantic information.

%\cite{Yangcao:2011} uses the positions and orientations of edgels (detected edge pixels) as feature to encode the input sketch and images.

%\paragraph{Multi-scale Descriptor}
%\cite{Daoudi:2000} matches sketch to images of object contours using a feature called curvature scale space (CSS), where the spatial relationship among contours are described using the relationship among their bounding boxes.
%\cite{Hu:2010} uses multi-scale HOG where HOG features are extracted in image regions of different sizes.

%\ryn{Briefly summarize some sketch-based retrieval systems here: \cite{Funkhouser:2003} and \cite{shao2011discriminative}.}
%
%\cite{shao2011discriminative} repeatedly transform and match.

\textbf{Part-based Models.}
In recent years, part-based models have been widely used in the computer vision community for the detection or recognition of objects in images. For example,
Felzenszwalb and Huttenlocher \shortcite{FelzenszwalbH05} present a pictorial structure model to encode the relationship among different body parts. More recently, Felzenszwalb et al. \shortcite{FelzenszwalbGMR10} introduce the Deformable Part Model (DPM), which is able to successfully identify complex objects. {Ferrari et al. \shortcite{FerrariFJS08} proposes a $k$AS feature for object detection, with each ``part'' constructed by linking $k$ roughly-straight adjacent contour segments.}
%These methods train part-based object templates and use them to detect objects, which demonstrate the usefulness of the features of parts.
However, it is unclear how to apply these models designed for object detection to our shape retrieval problem.
%DPM uses root detector to find a match of the whole object, and then uses part detectors to fine-tune the result.
% The feature is about ``parts'' existing in the image, where each ``part'' is constructed by linking $k$ roughly-straight adjacent contour segments. Our proposed \featname~feature differs from $k$AS in two ways: 1) \featname~encodes meaningful (functional) parts of objects, and 2) it encodes them in multiple scales.

\textbf{Image Pyramids.} Since the pioneering works in~\cite{burt1981fast,Burt:1983}, pyramid methods have been extensively used for image analysis to capture the underlying patterns in multiple scales.
For example, Lazebnik et al. \shortcite{Lazebnik2006} introduce the idea of  spatial pyramid matching (SPM) for natural scene categorization. SPM uses features extracted in regions of different sizes, and organizes them into a spatial pyramid. This idea has been later extended and used in many applications, such as image classifications~\cite{yang2009linear}, image matching~\cite{Shrivastava:2011} and 3D object recognition~\cite{li20073d,Redondo:2012}. We also adopt this idea of spatial pyramid, since it is proven to be more effective than single-level approaches. However, unlike existing methods, which construct pyramids of pixels, our method constructs a pyramid of semantic parts.

%\james{suggest remove this part}
%\textbf{Partial Matching.}
%Partial matching has been well-studied in example-based 3D retrieval. It uses a 3D shape as the query and returns 3D objects that contain similar parts with the query shape~\cite{funkhouser2006partial,suzuki2005partial,tabia2011new,tierny2009partial}.
%Existing methods for partial matching are mainly based on two techniques:
%1) utilizing feature descriptors that support partial matching, such as \ryn{(*** give some example features. ***)} ~\cite{funkhouser2006partial,tabia2011new} and
%2) employing the decompose-and-match strategy by first decomposing the models into components and then measuring the similarity between sets of components.
%%
%There are also a few partial matching methods that use 2D sketches as input.
%\cite{Eitz:2012:SketchRetrieval} proposed the Bag-of-Words model that inherently supports partial matching. However, we argue that comparing the feature of the whole sketch with the feature of the a 3D part is not optimal, and our proposed representation facilitates comparison between parts of a sketch and parts of a 3D object, which is shown to be more effective.
%\cite{shao2011discriminative} proposed a method that repeatedly transforms and matches the sketched part to the database images. This approach has a high complexity. In contrast, our method only needs to execute several nearest-neighbor searches, after extracting the feature of the partial sketch in constant time. Hence, logarithmic time complexity is possible, resulting in a much better scalability.

\section{\featname}\label{sec:feature}

%\inshort{how to extract the \featname feature from a \emph{segmented line drawing}?}

We assume that
%an input sketch, which can be either a query sketch or 2D model contour,
both the input query sketch and the 2D model contours have been pre-segmented into semantic parts.
%We will discuss how to get semantic segmentation of sketches in Section~\ref{sec:framework}. (*** This semantic segmentation information is a key information and needs to be defined at the beginning. ***)}
Due to the multi-scale nature of objects, it is not uncommon that a query sketch and a model contour correspond to the same object but have different segmentations. It is thus important to know the scale of each part and compare parts only at the same scales.

Since each part does not have its corresponding label, it is challenging to form a semantically meaningful hierarchy of parts for matching. Instead, we adapt the idea of image pyramids into our problem for part scale normalization. Note that the query sketch and the model contours are both represented by pyramids of parts. The use of a common pyramid for both of them not only makes it possible to compare parts at the same scales but also capture the multi-scale nature of objects. Although the following discussion focuses mainly on how we process input sketchs, model contours are processing in exactly the same way.

\textbf{Definition.}
Like image pyramids, a \emph{\featname}~consists of multiple scale levels, with each level containing groups of pre-segmented parts in the input sketch. Each group of parts as a whole at the same level have similar scale, and upper levels have larger scales, as illustrated in Figure~\ref{fig:grouping}.

\subsection{Pyramid Generation}\label{sec:grouping}

\begin{figure*}[th!]
\begin{center}
	\includegraphics[width=0.9\linewidth]{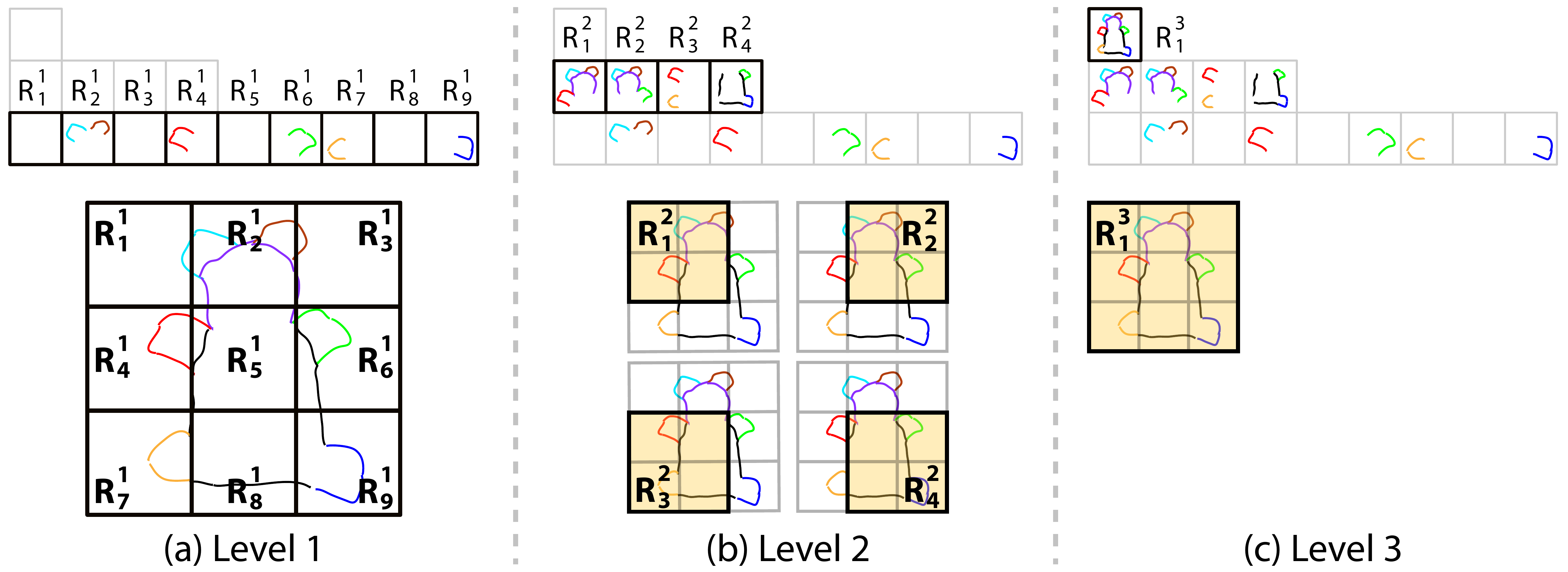}
\end{center}
	\caption{
		Constructing a 3-level \emph{\featname}~by assigning the semantic parts (color-coded) to different regions of the pyramid. $R^j_i$ is the i-th region in j-th level.}
	\label{fig:grouping}
\end{figure*}

%\inshort{how to generate the pyramids of parts?}

%\hongbo{Move this paragraph to a later section}
%To generate the pyramid of parts (Figure~\ref{fig:pipeline} (c)) given a line drawing, the atomic parts are required. If the line drawing is a user sketch, the user will be asked to provide the segmentation. If the line drawing comes from the rendering of a 3D model, the model must have been segmented, and the segmentation will be transferred to the line drawing (row 2 and 3 in Figure~\ref{fig:pipeline} (a)(b)).

To generate a \emph{\featname}, we associate each level with a set of regions in the sketch image. Let $R^l_i$ denote a region at level $l$. For example, $R^1_1$, $\ldots$, $R^1_9$ are the nine regions at level 1, as illustrated in Figure~\ref{fig:grouping}. Each region corresponds to a group of parts. Note that a region might be associated with zero (e.g., $R_1^1$), one (e.g., $R_4^1$) or multiple parts (e.g., $R_2^1$). Parts associated with a region $R^l_i$ is considered as one group of parts at level $l$. Since groups of parts at upper levels are of larger scale, we require larger regions at upper levels.

%A pyramid of parts is defined as groups of atomic parts, where each group is within a region in the image. The regions are organized into levels, where higher level has larger but fewer regions than any level below. A concrete example with 3 levels is shown in (Figure~\ref{fig:grouping}).

%The atomic parts will be grouped into a pyramid of several levels, as shown in Figure~\ref{fig:grouping}. Each level has several regions, each of which corresponds to a square area in the line drawing (Figure~\ref{fig:grouping} (a)), and the atomic parts will be grouped and assigned to these regions.

The main criteria to determine if a part $a$
%\hongbo{maybe use $P$?}
belongs to a region $R$ is to check whether $a$ is enclosed by $R$ or not. When $a$ is completely within $R$ (e.g., the leg part in $R_9^1$ in Figure~\ref{fig:grouping}), it is easy to conclude that $a$ should be in the group of parts associated with $R$. However, the problem becomes tricky when $a$ covers multiple regions (e.g., the red arm in Figure~\ref{fig:grouping} covering $R^1_4$ and $R^1_5$). Some part may cover multiple regions at the current level mainly because it should belong to a region at an upper level. For example, the body part in Figure~\ref{fig:grouping} belongs to region $R^2_4$, instead of $R^1_5$, $R^1_6$ or $R^1_8$.

With the above observations we determine if $a$ is assigned to $R$ by considering both the inclusion of $a$ inside $R$ and the relative size of $a$ to $R$.
%A straightforward approach is to assign an atomic part $a$ to a region $R$ if more than half of $a$ falls within $R$. However, as atomic parts usually go across multiple regions, the assignment is inherently uncertain, and some atomic part lying in the middle of two regions might get assigned to any one of them just because of some small error in the line drawing. Furthermore, user sketch is highly variable as parts get deformed or shifted, making the assignment unstable. Therefore, we take a conservative approach, considering both
Specifically, we formulate the likelihood of $a$ belonging to $R$ as $p(a)$:
\begin{equation}\label{eq:grouping}
p(a) = W_s(a)W_l(a),
\end{equation}
where $W_s(a)$ enforces penalty when the size of $a$, denoted as $S$ and defined as the longer side of the bounding box of $a$, is larger than $\beta L$. Here $\beta$ is a parameter ($\beta = 0.85$ in our implementation) and $L$ is the length of the longer size of $R$. Precisely, when $S \le \beta L$, we set $W_s(a)=1$, corresponding to no penalty. Otherwise, $W_s(a)$ decreases as $S$ deviates from $\beta L$, which is computed as $W_s(a)=1.1e^{\frac{-(S-\beta L)^2}{\sigma^2}}-0.1$, and $W_s(a)$ is clamped to 0 if it becomes negative.

Even if the size of $a$ is smaller than the size of $R$, it is still possible that $a$ extrudes from $R$, since $a$ is not necessarily centered at $R$. We thus use $W_l$ to penalize the extrusion of $a$ from $R$. Let $|a|$ be the stroke length of $a$, $a \cap R$ be the part of $a$ inside $R$, and we have $W_l(a) = |a \cap R|/|a|$, which reaches the maximum when $a$ is completely inside $R$.

%where
%\begin{align*}
%& S = max(w,h) \\
%& W_s(a) =
%	\begin{cases}
%		1, & \mbox{if } S \le \beta L \\
%		max(0,e^{\frac{-(S-\beta L)^2}{\sigma^2}}*1.1-0.1), & \mbox{otherwise} \\
%	\end{cases} \\
%& W_l(a) = |a \cap R|/|a|
%\end{align*}
%
%$\beta$ is set to 0.85 in our implementation. Here $W_s$ penalizes atomic parts whose sizes are larger than 85\% of the region size $L$, and $W_l$ penalizes extrusion from the region.

In the end, $a$ is assigned to $R$ if $p(a) \ge 0.5$.
Also, we compute a reliability value for each region to quantify the certainty of the assignments happened to this region, which in later stages is used to downplay those regions having many uncertain assignments of parts. Let $\{a_1,a_2,...,a_n\}$ be the parts assigned to region $R$, then the reliability of $R$ is computed as $c(R)=\sum w_ip(a_i)$, where $w_i=|a_i|/\sum |a_i|$. Each part will eventually get assigned to at least one region, because the topmost level region ($R^1_3$ in Figure~\ref{fig:grouping}) covers the entire image.

\subsection{Feature Extraction}\label{sec:gabor}

%\inshort{how to do gabor filtering and generate the final feature?}

\begin{figure}[h]
	\includegraphics[width=\linewidth]{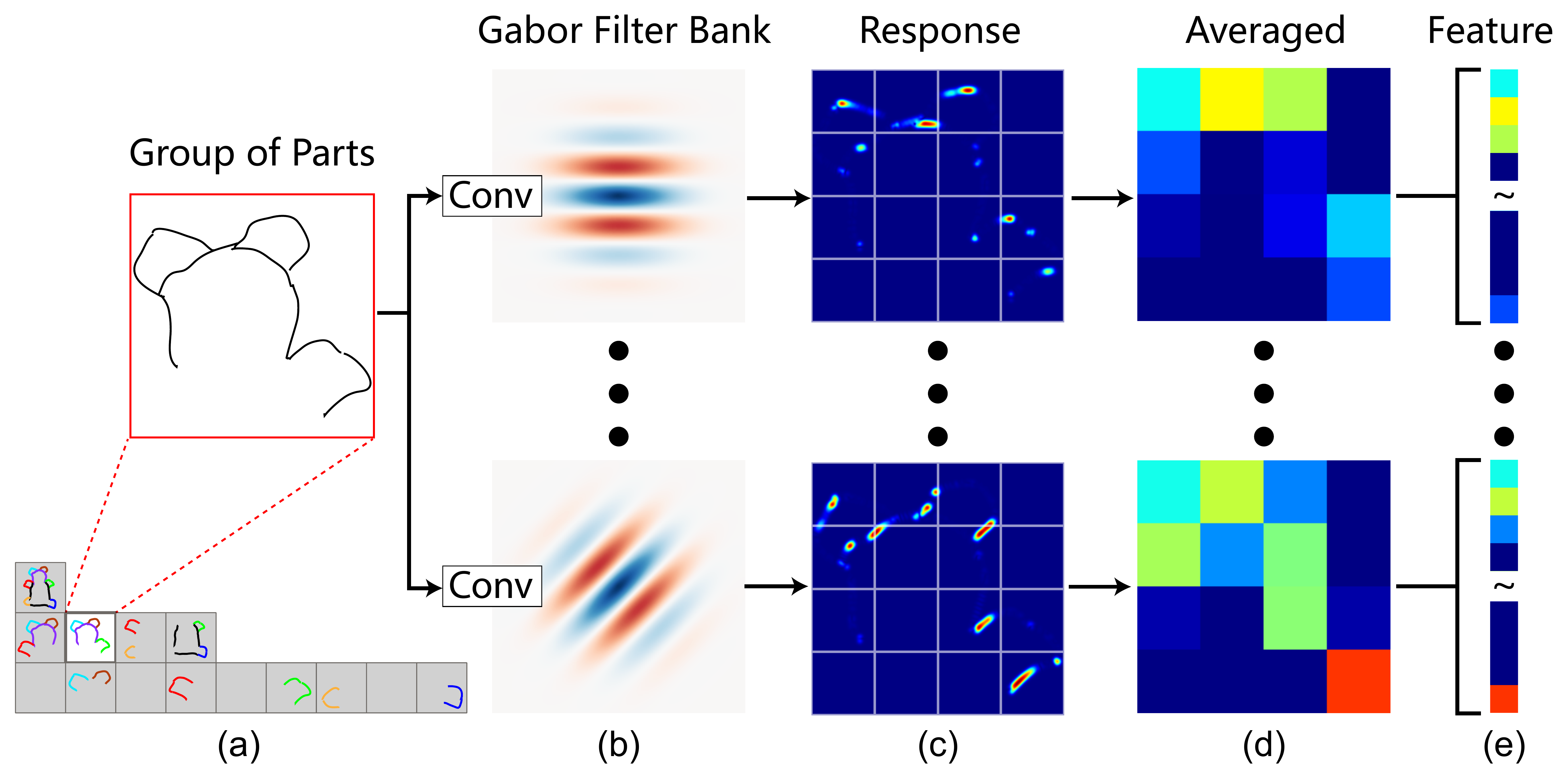}
	\caption{Extracting the feature for a group of parts with gabor filters. The group is first placed in a bounding square (a), and then convolved with a set of gabor filters (b), each resulting in a response map (c), which are further averaged by a coarse grid (d) and concatenated into the final feature (e).}
	\label{fig:gabor}
\end{figure}

The \emph{\featname}~feature is the concatenation of all the features extracted from all the regions. To begin with, each of the regions in the pyramid is either empty or contains a group of parts. For empty regions, their features are simply all zeros. For others, their features are gabor features extracted from the groups of parts in them, as shown in Figure~\ref{fig:gabor}. A group is first placed in a bounding square, and then convolved with a set of gabor filters. Each response is averaged by a grid (Figures~\ref{fig:gabor}(c) and~\ref{fig:gabor}(d)), and the outcome becomes part of the final feature (Figure~\ref{fig:gabor}(e)). The parameters of the gabor filters are different for each level of the pyramid, and is discussed in Section~\ref{sec:experimentalsetting}.

%The parameters of the gabor filters are as follows: peak response frequency $\omega_0=0.1$, frequency bandwidth $\sigma_x=3$, angular bandwidth $\sigma_y=9$, and the orientations $\Theta$ are $0,\pi/4,\pi/2,3\pi/4$. The explanation of these parameters can be found in~\cite{Eitz:2012:SketchRetrieval}. We have used a setting different from~\cite{Eitz:2012:SketchRetrieval} because the linewidth of our sketch is one pixel, rather than $\sigma_x/\omega_0$ in their work.

\section{Shape Retrieval Framework}\label{sec:framework}

Our shape retrieval engine is built upon the \emph{\featname}~feature, and the pipeline is shown in Figure~\ref{fig:pipeline}. As in \cite{Eitz:2012:SketchRetrieval}, we take a 2D-to-2D matching approach, i.e., matching the input sketch against all the views of all the models in database. To construct the database, we render each model for each selected view using suggestive contour~\cite{decarlo2003suggestive} (Figure~\ref{fig:pipeline}(b)), and extract its \emph{\featname}~feature (Figures~\ref{fig:pipeline}(d) and~\ref{fig:pipeline}(e)). Given a query sketch, its \emph{\featname}~feature will be extracted and matched against all the features in the database, after which the top matched models will be retrieved.

%\begin{figure*}[th!]
%	\includegraphics[width=\linewidth]{figure/pipeline}
%	\caption{Our shape retrieval pipeline. The first row shows the pipeline for processing the input sketch. The input sketch is pre-segmented into semantic parts (color-coded) (b), which are assigned to different image regions and then grouped into a pyramid (c). Gabor features are then extracted from each group of parts (d), which are concatenated into the \emph{\featname}~feature (e). The last two rows illustrate the processing of database models. A pre-segmented 3D model (a) is first rendered into a 2D model contour using Suggestive Contours (b), where segmentation is transferred from the model. After that, the process is similar to processing of the input sketch. Finally, the input sketch is matched with each contour (or view) of every model, and a ranked list of similar models is returned (f).}
%	\label{fig:pipeline}
%\end{figure*}

\subsection{The Query Sketch}
%\inshort{how to sketch and how to do segmentation?}

%\begin{figure}
%	\includegraphics[width=\linewidth]{figure/sketchseg}
%	\caption{\cq{need to delete this figure.}The input sketch and the segmentation.}
%	\label{fig:sketchseg}
%\end{figure}

The input query sketch consists of a set of strokes drawn by the user. The sketch is scaled such that a fixed-sized canvas (of resolution $320 \times 320$ in our implementation) forms its bounding square. To extract the \emph{\featname}~feature, segmentation of the sketch is required, which can be done automatically~\cite{Sun:sketchsegmentation2012,Huang:2014sketch} or manually~\cite{Noris:2012a}. For maximum accuracy, here we opt for the manual approach, which is discussed in detail in Section~\ref{sec:experimentalsetting}.
%\cq{[I suggest to remove these sentence and Fig.\ref{fig:sketchseg} since I will describe it in the "Tools paragraph" of the experiment settings.] For maximum accuracy, here we let the user segment the sketch manually by drawing zones with closed curves around the atomic parts (Figure~\ref{fig:sketchseg}, second row), and each stroke is assigned to the zone that contains more than half of it in terms of stroke length. Those strokes not assigned to any zone are assigned to a background zone. Finally, the strokes in each zone (including the background zone) constitute an atomic part.}

\subsection{Database Construction}

To construct a 3D database, we need a set of segmented 3D models. This database stores the \emph{\featname}~features of the 2D contours of each segmented models under a selected set of views.

The procedure of computing the \emph{\featname}~feature of a 3D model is shown in the second and third rows of Figure~\ref{fig:pipeline}. Given a segmented 3D model, a 2D model contour is generated from a given view of the model using Suggestive Contours~\cite{decarlo2003suggestive}, with the segmentation information transferred from the 3D model (Figure~\ref{fig:pipeline}(b)). The semantic parts are then processed into a \emph{\featname}~(Figure~\ref{fig:pipeline}(c)), which is used to produce the \emph{\featname}~feature (Figure~\ref{fig:pipeline}(d)) following the procedures described in Section~\ref{sec:feature}.
Multiple views are used for each model, which are representative views generated using~\cite{zou2014VAR}. The view generation process starts by sampling many views uniformly distributed on the viewpoint sphere, among which 42 views covering most of the information given by the dense views are selected.

\subsection{Retrieval}
%\inshort{how to compute feature distance?}

To retrieve a model, the \emph{\featname}~feature of the sketch is compared with all the \emph{\featname}~features in the database, and the $K$ nearest neighbors are returned as matches. The distance between two \emph{\featname}~features is the weighted sum of distances between the constituent gabor features in the corresponding regions.
Let ${\mathbf{x}=(x_1,x_2,...,x_{n})}$ be a \emph{\featname}~feature, where $x_i$ is the gabor feature of region $R_i$. (The level is not important here.) The distance of two features $\mathbf{x}$ and $\mathbf{y}$ is computed as:
\begin{equation}\label{eq:distance}
D(\mathbf{x},\mathbf{y})=\sum_{i=1}^{n}w_i||x_i-y_i||
\end{equation}
The weight $w_i$ is proportional to the product of region importance and reliability. For region $R_i$, the importance $m_i$ is set roughly proportional to the area of the region. In our 3-level implementation, and $m_i$ is set to 1, 4 and 9 for regions at levels 1, 2 and 3, respectively. The reliability $c(R_i)$ is the degree of certainty of assigning the semantic parts in $R_i$ to $R_i$, as described in Section~\ref{sec:grouping}. Given these quantities, the unnormalized weight $w_i'=m_ic(R_i)$, and the final weight $w_i=w_i'/\sum w_i'$.

\newcommand{\mdfull}{$\textbf{OUR-FULL}$}
\newcommand{\mdnogroup}{$\textbf{OUR-NOG}$}
\newcommand{\mdpixel}{$\textbf{OUR-PIX}$}
\newcommand{\mdstroke}{$\textbf{OUR-STK}$}
\newcommand{\mdglobal}{$\textbf{GF}$}
\newcommand{\mdbow}{$\textbf{BOW}$}

\section{Evaluation}
\label{sec:evaluation}
We have conducted four experiments, Exp.~1-4, to evaluate the performance of the proposed method.
Exp.~1 evaluates the performance when using different region subdivision strategies (Section~\ref{sec:regionsubdivision}).
Exp.~2 evaluates the performance when using user-provided sketch segmentation information (Section~\ref{sec:fullcompare}).
Exp.~3 evaluates the performance of a simple, automatic sketch segmentation strategy by grouping strokes (Section~\ref{sec:autoseg}).
Finally, Exp.~4 evaluates the performance on incomplete matching (Section~\ref{sec:incompletematching}).

\subsection{Experimental Settings}
\label{sec:experimentalsetting}

\textbf{Tools.} We have developed a prototype retrieval system, which we
used to collect input sketches and evaluate the performance of the proposed method.
The interface of the system allows users to draw three types of strokes: bounding box, segmentation and the query sketch. The user may draw these strokes in any order. The bounding box strokes represent a bounding box of the sketch, which is only useful for incomplete matching (Seciton~\ref{sec:incompletematching}), where a bounded canvas is needed. The segmentation strokes are used to segment the query sketch into semantic parts. Each segmentation stroke is a closed curve forming a \emph{zone}, and each stroke of a query sketch is assigned to the zone that contains more than half of it. All the query sketch strokes assigned to one zone are assumed to form one semantic part. Note that it is possible for one semantic part to lie completely within another (e.g., a human eye and head), and they cannot be separated because the zone for the larger semantic part will contain that of the smaller one. In this case, a query sketch stroke will be assigned to the smaller zone only if more than half of the stroke is inside it. Finally, if there exists $N$ semantic parts, only $N-1$ zones are needed, and the strokes not belonging to any zone are assigned to a background zone, which also represents one semantic part.

%First, it allows the user to input the query sketches. Second, it allows the user to specify a bounding box for the input sketch, as shown in Figure~\ref{fig:UI}(a). This function is only used in incomplete matching (Set \textbf{D}).
%Third, it allows the user to specify segmentation information.
%Our system treat each stroke in the second mode as the envelop
%of a part. The sketch points within one stroke in the second mode is integrated as a part.
%If one part is completely encircled in another part, the system removes the smaller part
%from the bigger one.

\textbf{3D models dataset. }Our 3D models come from the PSB dataset~\cite{Chen:2009}. This dataset contains 380 models in 19 categories. It contains segmentation results from different segmentation methods and we selected the segmentation results produced by Randomized Cut~\cite{Chen:2009}.

\textbf{Sketch dataset. }With our prototype retrieval system, we collected a total of 428 complete sketches and 205 partial sketches (see the supplemental).
Both full and partial sketches covered all 19 categories of the PSB model dataset.
10 users were invited to freely draw query sketches after we had shown them an example model from each category of the 3D dataset. Users were asked to freely specify the segmentation strokes for their drawings. This sketch dataset is used in most of the experiments where segmentation is needed. To compare with~\cite{Eitz:2012:SketchRetrieval} fairly when sketch segmentation is not available, we use a subset of their sketch data, which includes 395 sketches, covering 10 of the PSB model categories. Other sketches used by~\cite{Eitz:2012:SketchRetrieval} do not have a corresponding category in the PSB model dataset and thus are discarded.

%The input sketches for Exp. \textbf{3} are obtained from Eitz tt al.~\cite{Eitz:2012:SketchRetrieval}.
%395 sketches, which fall within 10 of the PSB model categories, are selected.

\textbf{Performance metrics.} To qualitatively evaluate the proposed method, we have adopted four performance metrics: 1) Precision-recall; 2) Top One (TO), which measures the precision of the top-one results, averaged over all queries;
3) First Tier (FT), which measures the precision of the top $N$ results
(where $N$ is the number of ground-truth models relevant
to the query), averaged over all queries; and 4) Mean Average
Precision (mAP), which summarizes the average precision of
ranking lists for all queries.

\textbf{Methods for Comparison.}
We mainly compared our framework with the popular Bag-of-Words framework (denoted as \mdbow)~\cite{Eitz:2012:SketchRetrieval}, and the global feature based framework (denoted as \mdglobal) as used in~\cite{eitz:2011}, which encodes the whole query sketch using a chosen shape descriptor (GALIF in our case). Our full method is denoted as \mdfull.
%It worth mentioning that the global feature based framework could obtain good retrieval results as evaluated in \cite{Eitz:2011}.
%For fair comparison, we utilize the GALIF descriptor in \cite{Eitz:2012:SketchRetrieval} to extract the features for the three frameworks compared.
As all the methods use Gabor filter somewhere, the parameters of the Gabor filters for all of them in all the experiments are set to the same (as described below).
%We have tried eight sets of "vocabulary" number ([800:100:1500]) for the $\textbf{BOW} framework, and found the retrieval performances on these eight sets of

\textbf{Parameters of the Gabor filters.}
Gabor filters are used in all the  methods compared, and the following parameters are shared among them: peak response frequency $\omega_0=0.1$, frequency bandwidth $\sigma_x=3$, angular bandwidth $\sigma_y=9$, and the orientations $\Theta$ are $0,\pi/6, \pi/3, \pi/2, 2\pi/3 ,5\pi/6$. The explanation of these parameters can be found in~\cite{Eitz:2012:SketchRetrieval}.
When averaging the Gabor response (Figure~\ref{fig:gabor}(d)), the grid size needs specified. For our method, it is 2x2, 4x4 and 6x6 for the image regions in Levels 1, 2 and 3, respectively. For \textbf{GF}, it is 6x6, same as the grid size used for the top-level image region in our method. For \textbf{BOW}, it is 4x4 as in~\cite{Eitz:2012:SketchRetrieval}.

%We use 4 tiles to subdivide the interested image patch (note: 4 tiles means the interested image area is subdivided into $4\times 4$ cells) in $\textbf{BOW}$ and the patches in Level 2 of our framework.
%We use 4 and 6 tiles for the patches in Level 1 and 3 of our framework, respectively.
%For $\textbf{GF}$, we use 6 tiles. We then obtain a $6\times4\times4 $ local feature vector for $\textbf{BOW}$,
%a $6 \times 6 \times 6$ global feature vector for $\textbf{GF}$, and
%a $9\times6\times4\times4 +4\times 6\times 4\times4+ 1\times6\times 4\times4 = 1344$ feature vector (on the group
%division strategy shown in Fig.~\ref{fig:grouping}) for the proposed framework.

\begin{figure}[h]
	\centering
	\includegraphics[width=0.9\linewidth]{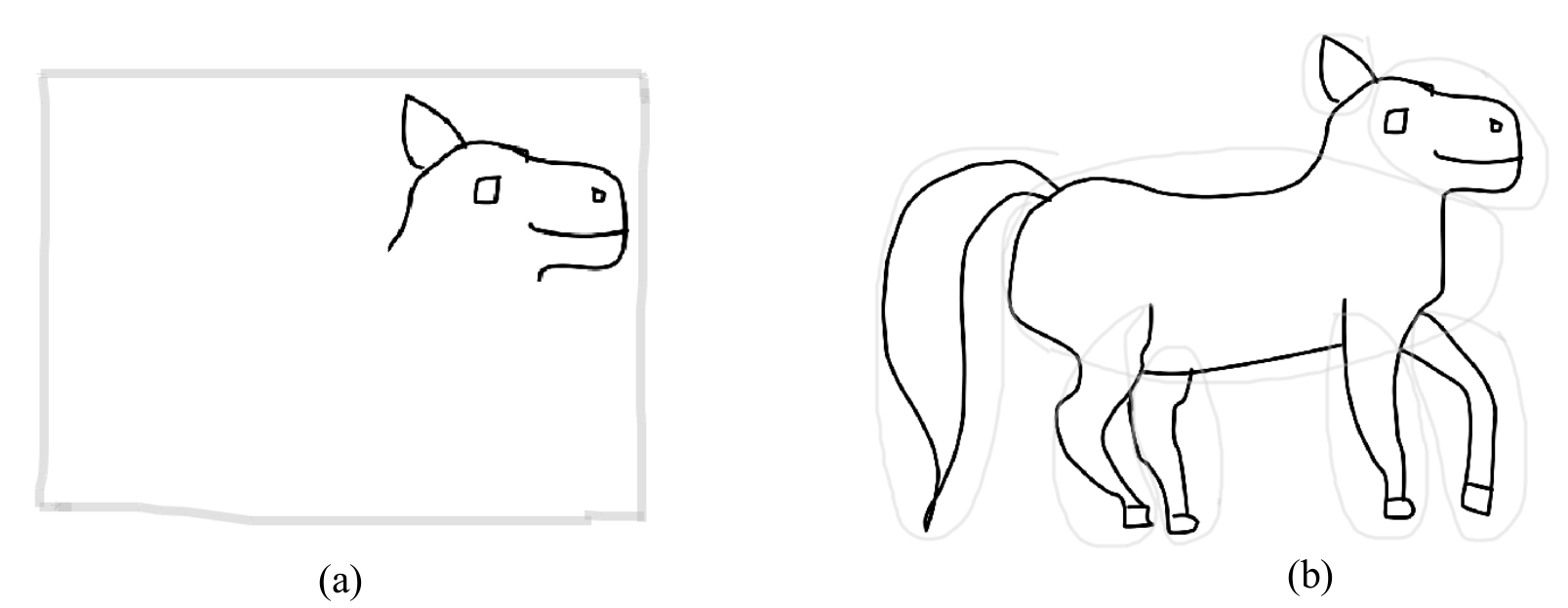}
	\caption{Three types of input strokes in our system: bounding box (left, gray), sketch (black) and segmentation strokes (right, gray).}
	\label{fig:UI}
\end{figure}

\subsection{Exp.~1: Region Subdivision Strategy}
\label{sec:regionsubdivision}
\begin{figure}[h]
	\centering
	\includegraphics[width=0.95\linewidth]{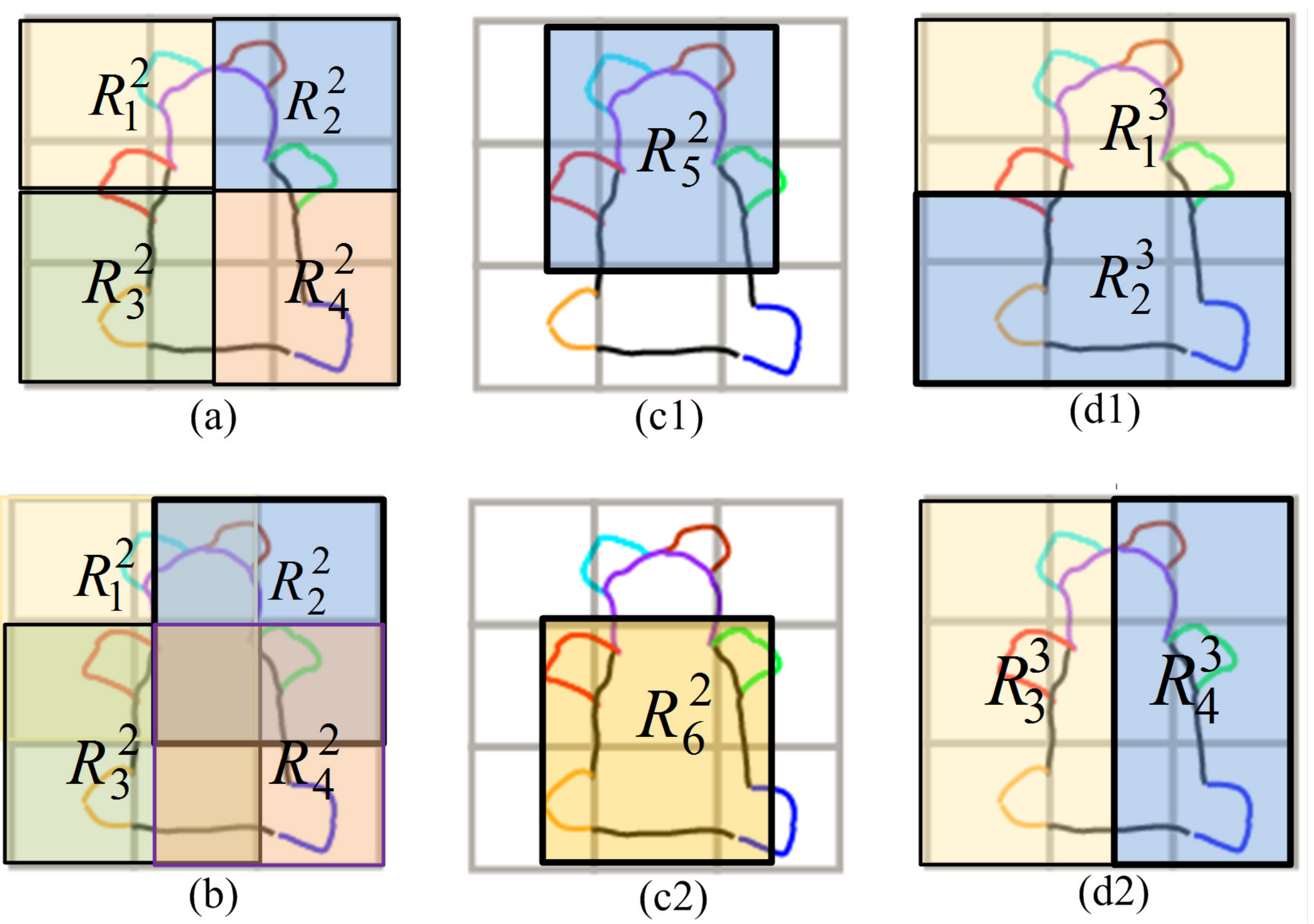}
	\caption{Different region subdivision schemes: (a) four regions without overlapping in Level 2 (i.e., $\textbf{4R\_NO}$); (b) four regions with overlapping in Level 2 (i.e., $\textbf{4R\_O}$); (c1)(c2) two new regions added to Level 2 (i.e., $\textbf{6R\_O}$); (d) regions of the new level added between Levels 2 and 3 of (i.e., $\textbf{4LV}$).}
	\label{fig:EvaGrouping}
\end{figure}

\begin{figure}[h]
	\centering
	\includegraphics[width=\linewidth]{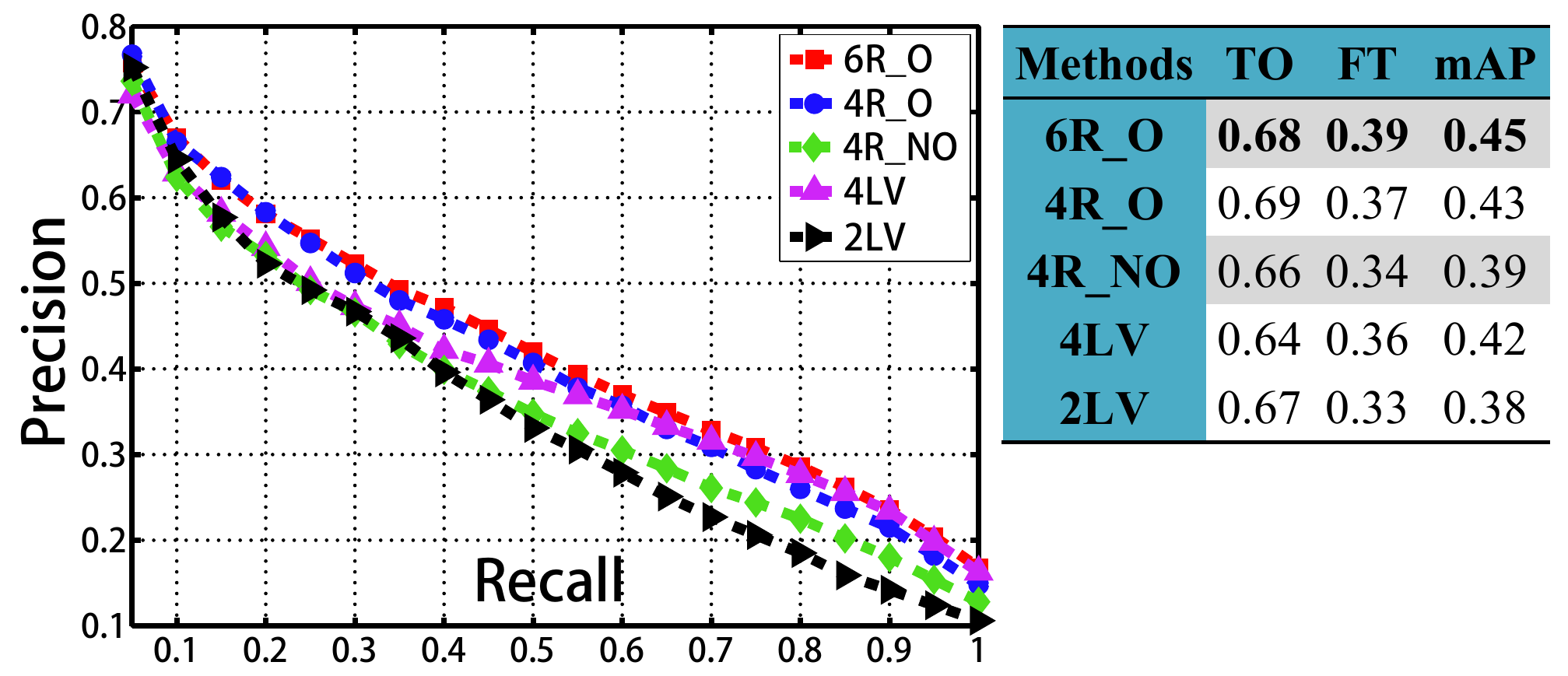}
	\caption{Performance comparison of different region subdivision schemes.}
	\label{fig:GroupingParamPeferm}
\end{figure}

Our method is based on \emph{\featname}~as illustrated in Fig.\ref{fig:pipeline}. The default number of levels for the \emph{\featname}~is 3 and the subdivision of regions is as shown in Figure~\ref{fig:grouping}.
In this experiment, we study the effect of using different subdivision schemes on our retrieval performance:
\begin{itemize}
\item Without overlapping regions: We divide the sketch into four $1.5L \times 1.5L$ regions, where $L$ is the side length of the square region $R_1^1$, as shown in Figure~\ref{fig:EvaGrouping}(a). This is denoted as $\textbf{4R\_NO}$.

\item Using different ways of constructing Level 2 regions: First, we divide the sketch into four different overlapped $2L \times 2L $ regions, as shown in Figure~\ref{fig:EvaGrouping}(b). This is denoted as $\textbf{4R\_O}$. Second, in addition to the original four regions shown in Figure~\ref{fig:grouping}(b), we add two new $2L \times 2L$ regions to Level 2 as shown in Figures~\ref{fig:EvaGrouping}(c1) and~\ref{fig:EvaGrouping}(c2). This is denoted as $\textbf{6R\_O}$.

\item Using a different number of levels: First, we add one more level between the current Levels 2 and 3 with four $3L \times 2L$ or $2L \times 3L$ regions to  $\textbf{4R\_NO}$. This is denoted as $\textbf{4LV}$. Second, we remove one level (Level 2) from $\textbf{4R\_NO}$. This is denoted as $\textbf{2LV}$.

\end{itemize}
In this experiment, the region subdivision for Levels 1 and 3 are fixed. Figure~\ref{fig:GroupingParamPeferm} compares the retrieval performances of the above four schemes. It shows that introducing region overlapping,
%($\textbf{4R\_O}$ vs. $\textbf{4R\_NO}$),
adding more regions
%($\textbf{6R\_O}$ vs. $\textbf{4R\_O}$)
and adding more levels
%($\textbf{4LV}$ vs. $\textbf{4R\_NO}$ vs. $\textbf{2LV}$, note that $\textbf{4R\_NO}$ has 3 levels)
all help improve the performance. This is because these operations increase the amount of information in the resulting feature, improving its discriminative power. Since the best performance is obtained using the scheme $\textbf{6R\_O}$, we use it in later experiments.

%and adding new regions in one level, and adding new levels will improve the retrieval accuracy.  By contrast, region overlapping and new regions in one level contribute more improvement than adding new levels.
%We ignore the truth that adding new levels on $\textbf{6R\_O}$ will bring more improvement on retrieval accuracy,
%and use $\textbf{6R\_O}$ as the region division rule in the next three sets of experiments.

\subsection{Exp.~2: Full Method Comparison}
\label{sec:fullcompare}

In this experiment, we compare our full method (\mdfull) to the two competing methods, namely Bag-of-Words model (\mdbow) and retrieval by global feature (\mdglobal), over the 428 segmented sketches collected using our system. The results are shown in Figure~\ref{fig:fullmethod} as red, purple and black curves, respectively.

%\begin{figure}[t]
%	\centering
%	\includegraphics[width=\linewidth]{figure/framework_performance1}
%	\caption{Precision-Recall curves and quantized performance based on TO, FT, and mAP on the 428 segmented sketches.}
%	\label{fig:RP1}
%\end{figure}

We can see that our method (\mdfull) has achieved the best retrieval performance on all four evaluation metrics.
\mdbow~has achieved the second best average retrieval precision (mAP), but its  retrieval accuracies evaluated by TO and FT are worse than those of \mdglobal.
These results motivated us to investigate if it is the multi-scale nature of the \emph{\featname} or the use of semantic parts that faciliates the better performance of \mdfull~over \mdbow~and \mdglobal.

\subsection{Multi-scale vs. Semantic Parts}\label{sec:factor}

Since our method adopts two main ideas, the multi-scale nature of the \emph{\featname} or the use of semantic parts, we would like to understand how two ideas contribute to the overall retrieval performance. Hence, we tested two approaches to evaluate the two ideas individually.

The first approach (denoted as \mdnogroup) skips the grouping stage, i.e., removing the effect of multi-scale. However, as the sketches may contain different number of semantic parts, if we simply extract a Gabor feature for each part, the final feature vectors will be of different lengths for different sketches, making them hard to compare. As such, to obtain a fixed-length feature vector, we still assign the semantic parts to image regions in one of the levels, but each semantic part is only assigned to one region (the one having the highest assignment likelihood in Eq.~\ref{eq:grouping}). After that, the process is the same as the full method.

The second approach (denoted as \mdpixel) removes all the information about semantic parts but keeps the multi-scale process. To do that, the sketch is rasterized into an image, and all segmentation information is discarded. In the grouping stage, the image patch bounded by each region is regarded as a part, which Gabor feature is extracted to compose the final feature.

\begin{figure}[t]
	\centering
	\includegraphics[width=\linewidth]{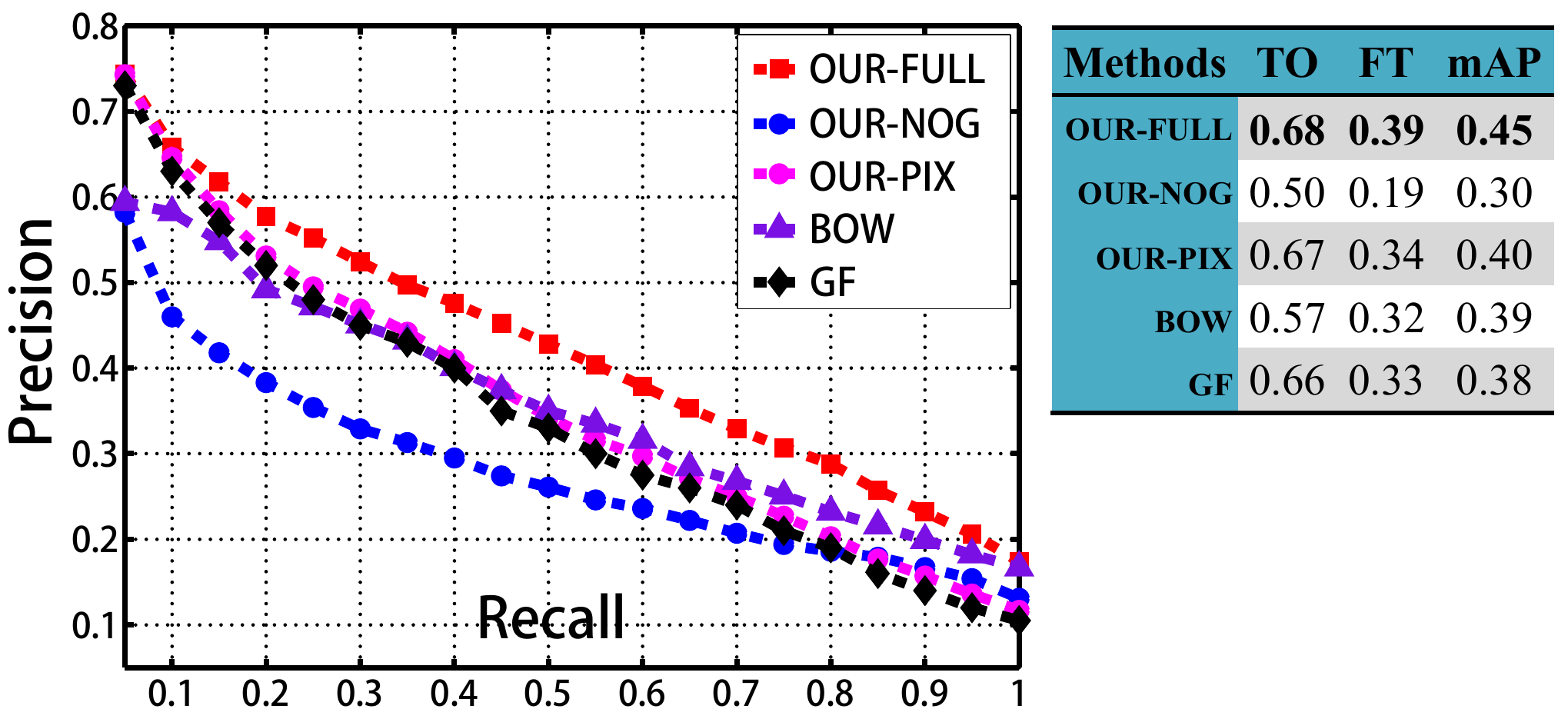}
	\caption{Retrieval performance on the 428 segmented sketches collected using our system.}
	\label{fig:fullmethod}
\end{figure}

The average retrieval performances of the two approaches on all the sketches are shown in Figure~\ref{fig:fullmethod}.
We can see that \mdpixel~performs only slightly better than \mdglobal, and \mdnogroup~performs much worst than \mdglobal~and \mdpixel, while \mdfull~performs the best.
This experiment shows that the combination of multi-scale and usage of semantic parts significantly improves the retrieval improvement than only using one of ideas.
From our analyses of the results, we have also found that \mdnogroup~often performs better on sketches that are segmented into a small number of parts by the user, such as the lower diagram shown in Figure~\ref{fig:motivation}(a). The main reason is that with a small number of parts, the segmentation of the 3D models tends to correspond to the segmentation of the input sketches.

%\subsection{Improving Inconsistent Segmentation with Multi-scale}\label{sec:group_or_not}

%\james{This experiment will show that grouping can help if the segmentation of the query sketch is not consistent with that of the database. If we do not have time, skip this.}

\begin{figure}[t]
	\centering
	\includegraphics[width=\linewidth]{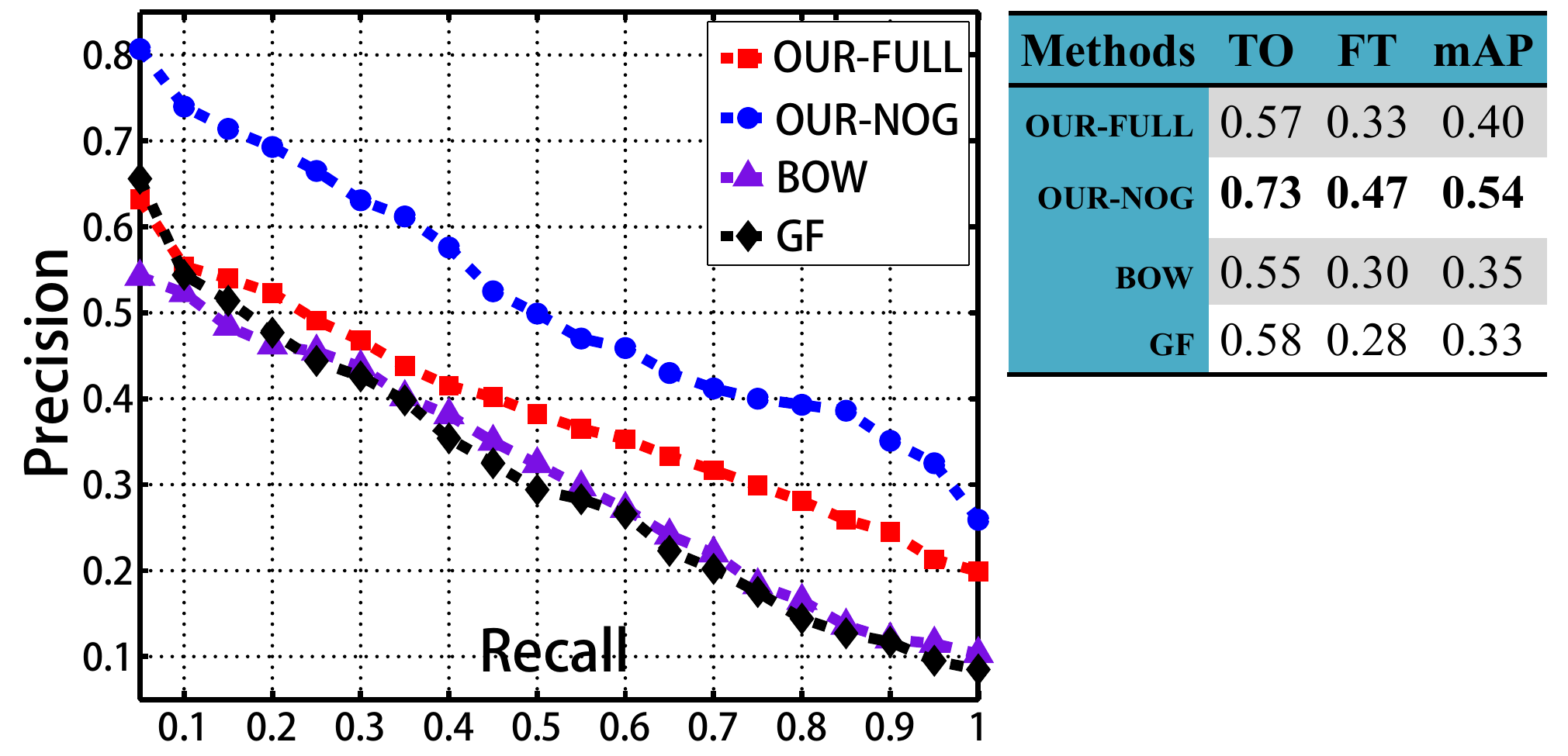}
	\caption{Retrieval performance on the 96 sketches having consistent segmentation information as the 3D models in the dataset.}
	\label{fig:RP2}
\end{figure}

To further evaluate this point, we selected all the sketches which segmentation information provided by the users are largely consistent with that of the 3D models in the dataset. There are 96 such sketches in total. Most of these sketches fall into three categories, "Cup", "Glass", and "Teddy Bear", where the segmentation is less ambiguous.
%and  results for both manual and automatic segmentation tools.
The retrieval results of these query sketches, as shown in Figure~\ref{fig:RP2},
indicate that \mdfull~significantly outperforms the other methods, when
the segmentation information of the input sketches is consistent with those of the  3D models.
%\ryn{(*** It may be good to state the inspiration of this observation here. ***)}

%inconsistent segmentation, no grouping vs grouping
%
%Figure~\ref{fig:missing}(\mdfull + \mdnogroup + \mdbow + \mdglobal~for inconsistently segmented sketches)

\subsection{Exp.~3: Strokes as Semantic Parts}\label{sec:autoseg}

\begin{figure}[h]
	\centering
	\includegraphics[width=\linewidth]{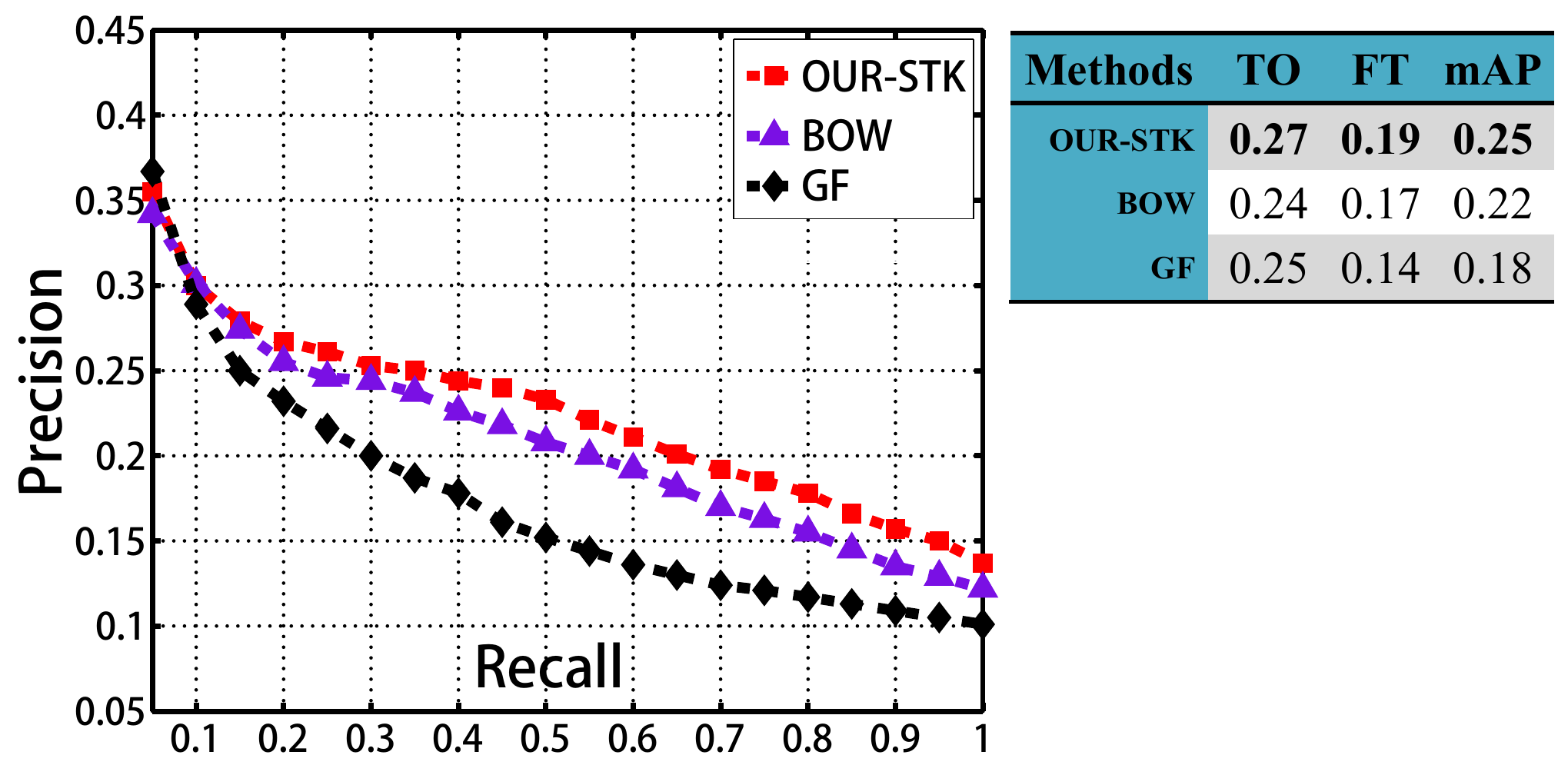}
	\caption{Retrieval performance using stroke-based segmentation on sketches provided by~\protect\cite{Eitz:2012:SketchRetrieval}.}
	\label{fig:PR3}
\end{figure}

In this experiment, we investigate how our method performs when segmentation information of the input query sketches is not available. A straightfoward approach to cope with this problem is to consider each stroke as a semantic part. This approach is denoted as \mdstroke~and is evaluated here over the sketch dataset provided by~\cite{Eitz:2012:SketchRetrieval}. The results are shown in Figure~\ref{fig:PR3}.

It is interesting to see that \mdstroke~performs better than \mdbow~and \mdglobal. The reason is that users' strokes tend to approximate the true segmentation to some extent.
These results also indicate that \mdfull~can achieve higher performance even on sketches without user segmentation.

\subsection{Exp.~4: Incomplete Matching}
\label{sec:incompletematching}

As the \emph{\featname}~feature is a collection of Gabor features obtained from different image regions, it is possible to compare the \emph{\featname}~features of some local regions only. This characteristic of the \emph{\featname}~feature suggests an interesting application, \emph{incomplete matching}, where a partially drawn query sketch can be used for model retrieval. Note that incomplete matching is not exactly the same as partial matching. With partial matching, the input sketch can be matched any local region of a database model. With incomplete matching, we may make use the location information of the drawn strokes relative to the canvas (or the user provided bounding box) so that the matching can be localized.

Here, it may be interesting to compare our method with \cite{Eitz:2012:SketchRetrieval}. Although \cite{Eitz:2012:SketchRetrieval} can also be used for incomplete matching, as their method computes some global statistics of local features, the comparison itself is therefore global, i.e., comparing the global statistics of an incomplete input sketch with those of the 2D model contour of a database model. On the contrary, with our method, we may simply skip the comparison of the Gabor features of those regions with no semantic parts in them.

\begin{figure}[h]
	\centering
	\includegraphics[width=\linewidth]{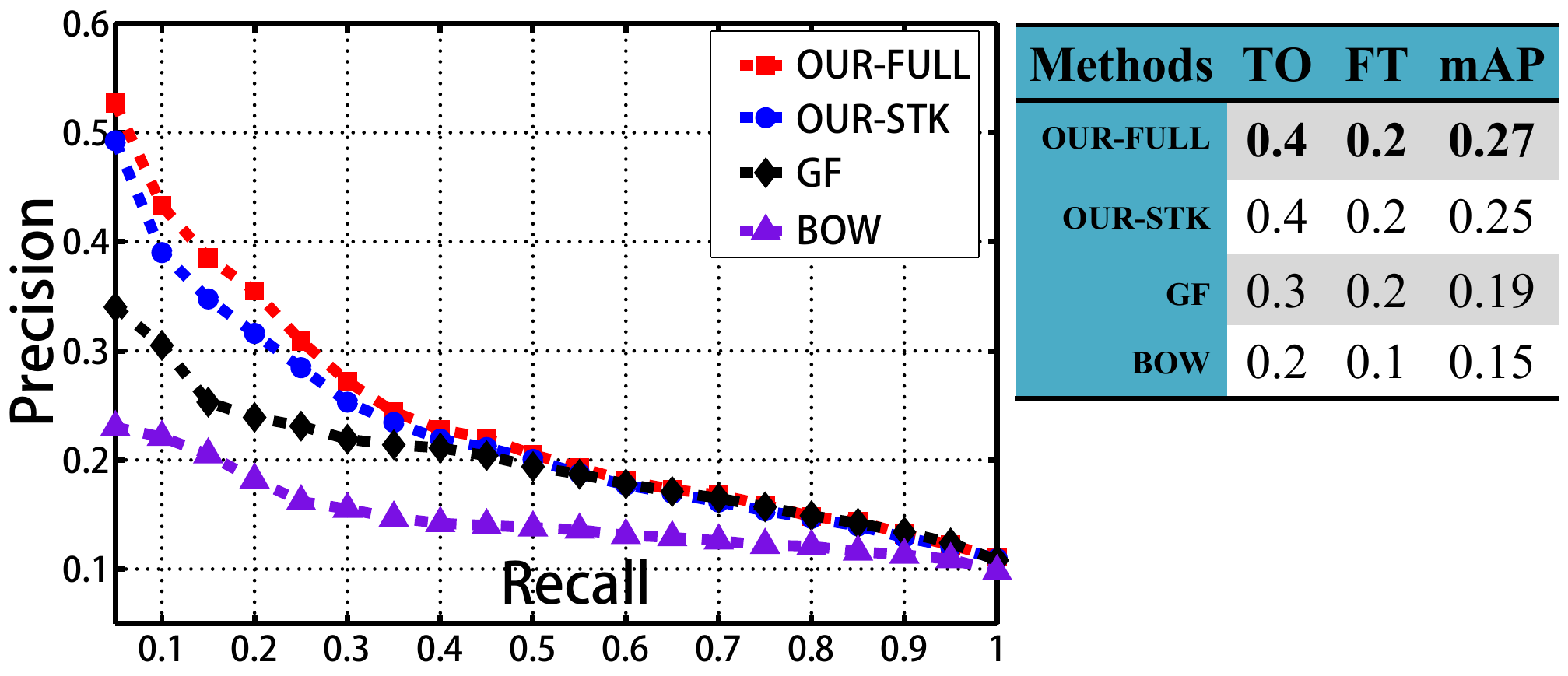}
	\caption{Retrieval performance of incomplete matching.}
	\label{fig:pmresult}
\end{figure}

To evaluate the performance of our method for incomplete matching, we have performed an experiment on incomplete matching. We collected 205 partial sketches covering all 19 categories of the PSB model dataset. The methods for comparison include our full method (i.e., \mdfull), our method without user segmentation but considering each stroke as a part (i.e., \mdstroke), Bag-of-Words model~\cite{Eitz:2012:SketchRetrieval} (i.e., \mdbow) and retrieval using global features (i.e., \mdglobal). Figure~\ref{fig:pmresult} compares the retrieval performances of the above methods. Our method outperforms the existing methods whether the segmentation information is obtained manually or automatically. This is mainly because the competing methods do not support localized matching, and they are matching the incomplete sketch to the complete model contours of the models.

%\cq{
%\mdbow only obtains sparse "words" on the mode of dense local patch extraction,
%and these less extracted words might be not enough to represent an incomplete 2-D shape.
%Therefore, it might generate senseless feature distances by comparing the histogram of the "word" distribution (usually sparse and short histogram columns) of an incomplete sketch and that of one model view (usually dense and tall histogram columns) in the database.
%It might explain the low performance of \mdbow when handling incomplete sketches. }
%
%\cq{\mdfull and \mdstroke outperform \mdglobal for the following two main facts:
%\begin{itemize}
%  \item \mdfull and \mdstroke only consider the feature distances within the interested regions for both
% model views and input sketch while \mdglobal takes into account "non-relevant" feature distances.
% For example, when the user have only drawn an incomplete sketch $\textbf{S}$ depicting a pair of legs of teddy bear
% as shown in $R_4^1$ and $R_6^1$ in Fig.~\ref{fig:grouping}, \mdfull or \mdstroke only computes the feature distances
% between the input sketch $\textbf{S}$ and model views $\textbf{V}$ on $R_4^1$ and $R_6^1$ while \mdglobal equivalent computes the feature distances
% between $\textbf{S}$ and $\textbf{V}$ on all the nine regions $R_{1,2,...,9}^1$.
%  \item \mdfull is invariant, to some extent, to the part translation and scaling (i.e., the feature of one part
% is constant as long as this part only translates and scales up or down within the same region), while
% \mdglobal is sensitive to the local translation or scaling of a part.
%\end{itemize}}

\section{Conclusion}

In this paper, we have investigated the use of semantic segmentation information to improve the performance of sketch-based 3D shape retrieval. We proposed the \emph{\featname} to support multi-scale matching of semantic parts. With the proposed method, we have evaluated the retrieval performances with and without user-provided segmentation information. Our experimental results show that the proposed method performs better than the state-of-the-art method by~\cite{Eitz:2012:SketchRetrieval} in both situations. We have also compared the two methods with incomplete input sketches. Our experimental results show that the proposed method performs significantly better than ~\cite{Eitz:2012:SketchRetrieval}.

%\bibliographystyle{acmsiggraph}
%\nocite{}
%\bibliography{paper}

\end{document}